\documentclass[twocolumn,showpacs,prx,amsmath,amstex,amssymb,citeautoscript,longbibliography]{revtex4-1}
\pdfoutput=1
\usepackage{natbib}
\usepackage[english]{babel}
\usepackage{letltxmacro}
\usepackage{latexsym}
\LetLtxMacro{\ORIGselectlanguage}{\selectlanguage}
\makeatletter
\DeclareRobustCommand{\selectlanguage}[1]{%
  \@ifundefined{alias@\string#1}
    {\ORIGselectlanguage{#1}}
    {\begingroup\edef\x{\endgroup
       \noexpand\ORIGselectlanguage{\@nameuse{alias@#1}}}\x}%
}
\newcommand{\definelanguagealias}[2]{%
  \@namedef{alias@#1}{#2}%
}
\makeatother
\usepackage[pdftex]{graphicx}
\usepackage{latexsym}
\usepackage{amsmath}
\usepackage{amssymb}
\usepackage{color}
\usepackage{color}
\usepackage{booktabs}
\usepackage{array}
\usepackage{hyperref}
\newcolumntype{C}{>{\centering\arraybackslash}p{1in}}
\AtBeginDocument{
\heavyrulewidth=.08em
\lightrulewidth=.05em
\cmidrulewidth=.03em
\belowrulesep=.65ex
\belowbottomsep=0pt
\aboverulesep=.4ex
\abovetopsep=0pt
\cmidrulesep=\doublerulesep
\cmidrulekern=.5em
\defaultaddspace=.5em
}

\newcommand{\e}{\varepsilon}

\renewcommand{\>}{\rangle}
\renewcommand{\(}{\left(}
\renewcommand{\)}{\right)}
\renewcommand{\[}{\left[}
\renewcommand{\]}{\right]}
\renewcommand{\v}[1]{\mathbf{#1}} 

\newcommand{\bs}[1]{\boldsymbol{#1}}
\renewcommand{\d}{\partial}

\newcommand{\eps}{\epsilon}

\newcommand{\header}[1]{\vspace{4pt}\noindent{\bf #1 --}}

\begin{document}
\title{Thermoelectric transport signatures of Dirac composite fermions in the half-filled Landau level} 
\author{Andrew C Potter$^1$}
\author{Maksym Serbyn$^1$}
\author{Ashvin Vishwanath$^{1,2}$}
\affiliation{$^1$Department of Physics, University of California, Berkeley, CA 94720, USA}
\affiliation{$^2$Department of Physics, Harvard University, Cambridge, Massachusetts 02138, USA}

\begin{abstract}
The half filled Landau level is expected to be approximately particle-hole symmetric, which requires an extension of the Halperin-Lee-Read (HLR) theory of the compressible state observed at this filling. Recent work indicates that, when particle-hole symmetry is preserved, the composite Fermions experience a quantized $\pi$-Berry phase upon winding around the composite Fermi-surface, analogous to Dirac fermions at the surface of a 3D topological insulator. In contrast, the effective low energy theory of the composite fermion liquid originally proposed by HLR  lacks particle-hole symmetry and has vanishing Berry phase. In this paper, we explain how thermoelectric transport measurements can be used to  test the Dirac nature of the composite Fermions by quantitatively extracting this Berry phase. First we point out that longitudinal thermopower (Seebeck effect) is non-vanishing due to the  unusual nature of particle hole symmetry in this context and is not sensitive to the Berry phase. In contrast, we find that off-diagonal thermopower (Nernst effect) is directly related to the topological structure of the composite Fermi surface, vanishing for zero Berry phase and taking its maximal value for $\pi$ Berry phase. In contrast, in purely electrical transport signatures the  Berry phase contributions appear as small corrections to a large background signal, making the Nernst effect a promising diagnostic of the Dirac nature of composite fermions.
\end{abstract}
\maketitle

In contrast to the incompressible fractional quantum Hall states at filling fractions $\nu=\frac{p}{2p+1}$, the 1/2-filled Landau level (LL) exhibits a compressible state with non-zero longitudinal conductance. The origin of this compressible state is naturally explained by the (CF) picture due to Halperin, Lee, and Read (HLR)~\cite{HLR}, in which electrons bind two flux quanta apiece becoming composite fermions~\cite{JainBook} that feel zero average orbital magnetic field and form a metallic state with a composite Fermi-surface. This composite Fermi liquid (CFL) state successfully captures many aspects of the experimental phenomenology of the $\nu=1/2$ state in GaAs. On the other hand (ignoring, for the moment, LL mixing), the 1/2-filled LL can be equivalently described by 1/2-filling an empty LL with electrons, or 1/2-depleting a full LL with holes~\cite{Kivelson97,Barkeshli15}. However, the HLR description begins by formally attaching flux to electrons (rather than holes) and naturally leads to a low energy description in terms of spinless CFs coupled to a Chern-Simons gauge field that breaks particle-hole symmetry (PHS) -- which we will henceforth refer to as the ``HLR state". While it is in principle possible that the experimental system chooses to spontaneously break PHS and form the HLR state, this option does not appear to be energetically favored in numerical simulations~\cite{Geraedts15}.

It was recently realized~\cite{Son15,Wang15short,Metlitski15,Wang15rev,MetlitskiSDuality,Mross15} that the HLR picture could be modified to produce a manifestly particle-hole symmetric candidate phase for the 1/2-filled Landau level -- the composite Dirac liquid (CDL). Like the HLR state, the CDL has a Fermi-surface of flux-like composite fermion excitations. However unlike the original HLR description in terms of spinless electrons bound to flux, these composite fermions are neutral Dirac-like particles that possess a pseudospin-1/2 degree of freedom, whose $x,y$ components are physically interpreted as an electrical dipole moment. This pseudospin is rigidly locked perpendicular to the CF momentum in close analogy to the Dirac surface state of a topological insulator (TI)~\cite{Son15,Wang15short,Metlitski15,Wang15rev}. The resulting pseudospin winding implies a Berry phase $\theta_{\rm B} = \pi$ whose value is fixed by PHS. On the face of it, the appearance of a single-Dirac cone in a purely 2D system with PHS appears to violate 
a fundamental fermion doubling theorem (valid also for interacting electron systems~\cite{Wang14,Wang14PRB}), and is ordinarily thought to only be possible at the surface of a 3D topological insulator. This apparent paradox can be resolved by noting that the action of PHS on the half-filled LL is inherently nonlocal, involving not only exchanging particles and holes, but also filling a topological LL band. Since this LL band has non-zero Chern number, it lacks a local Wannier basis and has at least one spatially extended orbital that carries the Hall conductance~\cite{Halperin82}. The action of filling or emptying this extended orbital is an inherently non-local, and this non-local action of PHS provides an interesting, and only recently realized~\cite{Son15,Wang15short,Metlitski15,Wang15rev}, exception to fermion doubling theorems, allowing certain kinds of topological-insulator surface state physics to be effectively realized in a 2D system.

In this paper, we examine thermoelectric transport signatures of the proposed Dirac-nature of CFs. While certain electrical transport signatures have been suggested~\cite{Son15}, for example examining the deviation of Hall conductance from ${e^2}/({2h})$, these proposals typically require accurately measuring small corrections to a large background signal, and it is desirable to look for more sensitive tests. To this end, we investigate the thermoelectric response for composite Fermi-liquid phases for the half-filled Landau level. 

Whereas longitudinal thermopower typically vanishes in a system with local PHS, we find that, due to the non-local realization of PHS in the half-filled LL, the Seebeck coefficient obtains a similar non-zero value in both particle-hole symmetric composite Dirac liquid and the particle-hole asymmetric HLR state, and does not qualitatively distinguish these two states. In contrast, we find that the transverse Seebeck coefficient (or closely related Nernst coefficient), provides a more sensitive probe for the degree of particle-hole asymmetry. Namely, in clean systems, this quantity is sensitive to the Berry phase of the Fermi-surface, vanishes in the maximally particle-hole asymmetric CFL state without Berry phase, and achieves a maximum value for the particle-hole-symmetric composite Dirac liquid. We show that a non-zero Nernst coefficient directly results from the composite Fermi-surface Berry phase and represents a qualitative signature of the composite Fermi surface topology. Moreover, we show how the Nernst effect in combination with conductivity and thermopower measurements can be used to quantitatively extract the composite Fermi-surface Berry phase. 

\header{Thermoelectric coefficients and PH Symmetry}
We begin by briefly recounting symmetry constraints on thermoelectric transport coefficients. The thermal drift of charged particles down a thermal gradient, $-\nabla T$, produces an electric current, such that in the presence of voltage and thermal gradients the electrical current can be written as $\v{j} = \hat\sigma\v{E}+\hat\alpha(-\nabla T)$, where $\hat\sigma, \hat\alpha$ are the electric and thermoelectric conductivity tensors respectively. Thermoelectric responses are typically measured in the absence of conducting leads, so that no net electric current flows through any cross section of the sample. In this geometry, an electric field $\v{E} = \hat S\nabla T$, must develop to cancel the thermally generated current, where, $\hat S=\hat\sigma^{-1}\hat\alpha$ is the Seebeck tensor, whose diagonal component $S^{xx}$ is typically referred to as simply, thermopower, and whose antisymmetric off-diagonal component $S^{xy}$ is often expressed through the closely related Nernst coefficient: $\nu_N = S^{xy}/B$. 

The longitudinal electric current, $\alpha^{xx}\(-\d_x T\)$, associated with charge-carrier flow induced by a thermal gradient has opposite signs for systems with electron and hole carriers, as carriers flow from hot towards cold regions independent of their charge. Hence {\it local} PHS, generated by exchanging particles and holes, $c(\v{r})\rightarrow c^\dagger(\v{r})$, constrains $\alpha^{xx},\sigma^{xy}=0$, implying vanishing thermopower, $S^{xx}=0$. In contrast, a number of thermopower measurements have been previously performed on the composite Fermi liquid state of the half-filled Landau level in GaAs~\cite{Ying94,Bayot95}. There, a sizable Seebeck coefficient was observed, whose magnitude agrees reasonably well with that of an ordinary Fermi liquid with the same density of particles as the composite Fermi-surface~\cite{Cooper97}. Given the expectation that particle-hole symmetry results in vanishing thermopower, such observations naively seem to rule out a (even approximately) particle-hole symmetric description of the CFL in GaAs.

Can one reconcile the experimentally observed large thermopower with the naive expectation of ordinary particle-hole symmetry? The key to resolving this apparent contradiction, is to note that the PHS present in the half-filled LL is inherently non-local, involving not only exchange of particles and holes but also filling a LL $|0\>\rightarrow \prod_{n\in LL} c_n^\dagger|0\>$, where $c_n^\dagger$ create an electron in the $n^\text{th}$ orbital of the LL. For example this non-local PHS exchanges $\sigma^{xy}\rightarrow -\sigma^{xy}+1\frac{e^2}{h}$ requiring non-vanishing Hall conductance: $\sigma^{xy}=\frac{1}{2}\frac{e^2}{h}$ in a PHS state. Then, non-vanishing thermopower can arise in a system with non-local PHS, due to the non-vanishing combination $S^{xx}=\rho^{xy}\alpha^{xy}$, where $\hat{\rho}\equiv\hat\sigma^{-1}$ is the resistivity tensor. We note that non-local PHS still constrains $\alpha^{xx}=0$, as a filled Landau level makes no contribution to $\hat\alpha$. Given these basic symmetry constraints on transport coefficients, we now turn to their computation from an effective field theory.

\begin{figure}[t!]
\includegraphics[width=0.8\columnwidth]{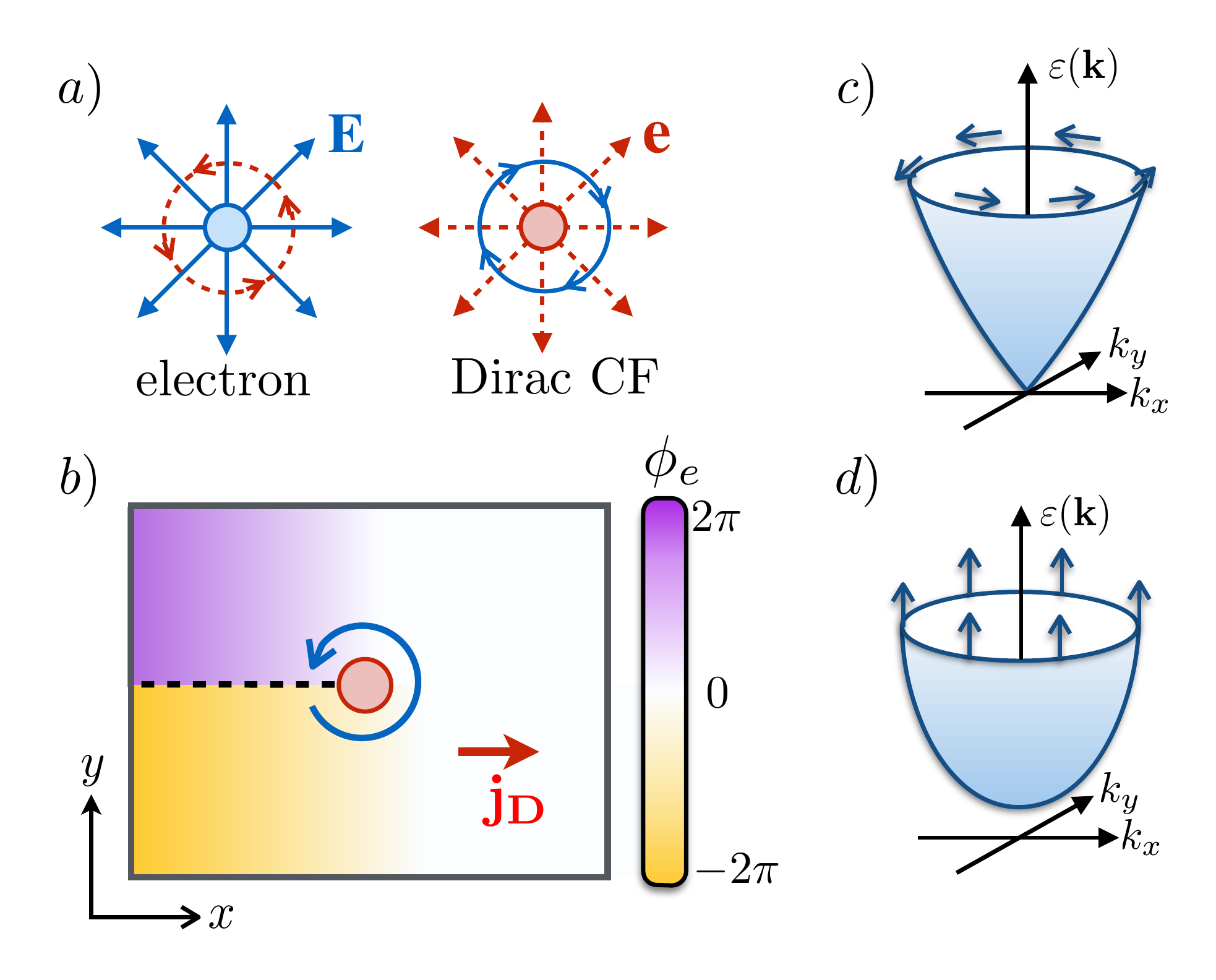}\vspace{-.1in}
\caption{ {\bf Schematic of charge vortex duality for transport - } a) An electron (blue dot) is a source of electric field $\v{E}$ (blue line) and $-4\pi$ flux of the emergent gauge field $b$ (circulating dashed red line indicates winding of phase for Dirac composite fermion). The Dirac composite fermion object (red dot) is a dual object that is a source of emergent electric field lines $\v{e}$ (dashed red lines) and $4\pi$ flux of the electron phase (circulating blue line).  Panel b) shows the electron phase, $\phi_e$, winding by $4\pi$ across the $y$-direction for each composite fermion propagating along $x$. Panels c,d) schematically depict the composite fermion dispersion and Fermi-surface pseudospin texture for the Dirac ($\pi$ Berry phase) and HLR ($0$ Berry phase) respectively.}
\label{fig:Duality}
\vspace{-.1in}
\end{figure}

\header{Thermoelectric properties of the composite Dirac liquid}
The composite Dirac liquid is described by the effective field theory~\cite{Son15}:
\begin{align}
\mathcal{L}_\text{eff} =& \psi^\dagger\(D_\tau-\mu-iv\v{D}\times \bs\sigma-m\sigma^z\)\psi+
\nonumber\\
&+\frac{iadA}{4\pi}-\frac{iAdA}{8\pi}+\int_{\v{r'}} \frac{b(\v{r}) b(\v{r}')}{4\pi^2\eps_r|\v{r}-\v{r}'|}+\dots
\label{eq:Leff}
\end{align}
where $\psi$ is the Dirac CF field, $\mu=\tau,x,y$ are (Euclidean) space-time indices, $a_\mu$ is an emergent U(1) gauge field whose magnetic field $b = \nabla\times a$ is $2\pi$ times the electron density, $D_\mu = \d_\mu + ia_\mu$ are covariant derivatives, and we have abbreviated Chern-Simons terms $\eps^{\mu\nu\lambda}A_\mu\d_\nu A_\lambda\equiv AdA$. The last term represents Coulomb interactions between electrons with dielectric constant $\eps_r$, and $(\dots)$ indicates other terms that are less important at low-energies. Particle-hole symmetry requires vanishing Dirac mass,~$m=0$. 

A straightforward derivation of Eq.~(\ref{eq:Leff}) from a parton description, as well as a computation of its transport properties obtained by a self-consistent RPA treatment of gauge fluctuations (in a similar spirit to analysis of conductivity in a U(1) spin liquid by Ioffe and Larkin~\cite{IoffeLarkin}) is presented in Appendices~\ref{app:Partons},\ref{app:PartonResponse}. Here, we instead obtain the same results by a simple physical picture by making an analogy to vortex motion in a superfluid. Specifically, due to the $\frac{adA}{4\pi}$ term in Eq.~(\ref{eq:Leff}), electrons see the Dirac composite fermions as vortices with flux $4\pi$ (in units $\hbar,c,e\equiv1$). The external magnetic field $B$ required to produce half-filling for the electrons, induces a finite density (equal to the electron density) of these vortices, (analogous to a superfluid subjected to $B>H_{c1}$). Just as for a superfluid, the fermionic vortices are non-local objects with long-range interactions which can be viewed as being mediated by a fluctuating emergent electromagnetic-like gauge field with vector potential $a^\mu$, whose electric charges are the Dirac composite fermions. However, unlike familiar bosonic vortices of a superconductor, these Dirac vortices have fermionic exchange statistics, and form a metallic state with a Fermi-surface. 

To examine the thermoelectric properties of this ``vortex metal" consider, a rectangular sample with spatial dimensions $L_x\times L_y$. A temperature gradient along $x$ tends to produce a Dirac CF current $j_D$, also along $x$. Each time a CF passes through, say, $x=0$, the electrons at $(x=0,y=L_y)$ acquire a relative $4\pi$ phase shift to those at $(x=0,y=0)$, resulting in a steady time-dependent winding of the electron phase-difference across the sample in the $y$ direction: $\Delta \phi_y(t) = 4\pi j_\text{D} L_y t$~(Fig.~\ref{fig:Duality}b). Such a time-dependent phase winding for electrons is gauge equivalent to a constant voltage difference across the $y$-direction of the sample: $V_y =- \frac{\d \Delta \phi_y}{\d t}$, corresponding to an electric field $E_y = \frac{V_y}{L_y} = -4\pi j_\text{D}$. In general, a composite fermion current, $\v{j}_\text{D}$ induces an electric field $\v{E}$:
\begin{align}
\v{j}_\text{D} = \frac{1}{4\pi}\hat{\eps}\v{E}
\label{eq:jD}
\end{align}
where $\hat\eps$ is the unit antisymmetric tensor with components $\eps^{ii}=0$, $\eps^{xy}=-\eps^{yx}=1$. Alternatively, this relation can be directly obtained by varying Eq.~(\ref{eq:Leff}) with respect to $\v{a}$ and setting the result to zero as demanded by the equations of motion. Note that, here and throughout the following discussion, unless explicitly stated otherise, we fix the magnetic field to be in the $+z$ direction. Expressions for the oppositely directed field can be obtained from our formulas by changing $\eps\rightarrow -\eps$.

Just as electrons see Dirac CF's as $4\pi$ flux, Dirac CF's see electrons as $-4\pi$ flux (see Fig.~\ref{fig:Duality} inset), and repeating the above arguments for a flow of electrons rather than CF's shows that an electrical current produces a transverse emergent electric field $\v{e}=4\pi \hat\eps\v{j}_{el}$. In addition, due to the half-integer Hall conductance ($AdA$ term in Eq.~(\ref{eq:Leff})) $\v{j}_{el}$ also induces an physical electric field $\v{E}=-4\pi\hat\eps \v{j}_{el}$. Together these imply:
\begin{align}
\v{j}_{el} = \frac{1}{4\pi}\hat\eps\(\v{E}-\v{e}\)
\label{eq:je}
\end{align} 
Since the electron system has both longitudinal and Hall conductance, we expect this electric field along $y$ to be accompanied by one along $x$, indicating non-vanishing thermopower coefficients, $\alpha_{xx},\alpha_{xy}$. 

To see how this works in detail we must understand how composite Dirac fermion currents $\v{j}_\text{D}$ are generated. In the linear-response regime, the Fermi-surface of CF's responds to a emergent $\v{e}$-fields and temperature gradients as:
\begin{align}
\v{j}_\text{D} = \hat\sigma_\text{D} \v{e}-\hat\alpha_\text{D}\nabla T
\label{eq:cfresponse}
\end{align}
We note in passing that the long-range current-current interactions between CF's produce inelastic scattering at a rate comparable to the energy of the CF quasiparticles relative to the Fermi energy. However, for the low-temperature DC thermopower responses, which are the focus of this paper, such inelastic scattering processes are less important than ordinary elastic scattering from impurities and can be neglected. 

Let us define:
\begin{equation}
\sigma_0 = \frac{k_F\ell}{4\pi};\,\,\,  \alpha_0= \frac{{\L} T}{v_F} \frac{\d\sigma_0}{\d k_F}
\label{definition}
\end{equation}
 where $\L = \frac{\pi^2k_B^2}{3e^2}$ is the Lorenz number, and $v_F,k_F$ are the composite Fermi velocity and wave-vector respectively, in units where $e,\hbar=1$. 
In this regime, $\hat\sigma_\text{D} \approx \sigma_0 \hat{1}$  is the DC conductivity of the composite Fermi surface with Fermi-wave-vector $k_F$ and transport mean-free path $\ell$, and similarly $\hat\alpha_\text{D}\approx \alpha_0\hat 1$ is the composite Fermi-surface thermoelectric conductivity. The transverse (Hall) components $\sigma^{xy}_\text{D},\alpha^{xy}_\text{D}$ vanish by PHS (which acts as an effective time-reversal symmetry on the CFs~\cite{Son15,Wang15short,Metlitski15}). While impurities locally break PHS, and only preserve PHS on average, for the long-wavelength impurity potentials relevant to GaAs, we can show that such extrinsic sources of $\sigma^{xy}_D$ are negligibly small (see Appendix~\ref{app:Disorder}). Lastly, we note that the response coefficients $\sigma_D,\alpha_D$ should be interpreted as transport coefficients (i.e. obtained from Kubo formulae by appropriately subtracting bound magnetization current contributions, see Appendix~\ref{app:PartonResponse}). 

\renewcommand{\arraystretch}{1.3}
\begin{table*}
\begin{tabular}{lCCC CCC}
\toprule
         & $\sigma^{xx}$ & $\sigma^{xy}$& $\alpha^{xx}$ & $\alpha^{xy}$& $S^{xx}$ & $S^{xy}$\\ \midrule%
HLR & $\dfrac{\sigma_0}{1+\(4\pi\sigma_0\)^2}$ &$\dfrac{1}{4\pi}\dfrac{(4\pi\sigma_0)^2}{1+(4\pi\sigma_0)^2}$ & $\dfrac{\alpha_0}{1+(4\pi\sigma_0)^2}$&$\dfrac{4\pi\sigma_0\alpha_0}{1+(4\pi\sigma_0)^2}$&$\sigma_0^{-1}\alpha_0$&$0$\\%
Dirac& $\dfrac{1}{\(4\pi\)^2\sigma_0}$ & $\dfrac{1}{4\pi}$ & $0$ & $\dfrac{\alpha_0}{4\pi \sigma_0}$ & $\dfrac{(4\pi)^2\sigma_0\alpha_0}{1+(4\pi\sigma_0)^2}$ & $ \dfrac{4\pi\alpha_0}{1+(4\pi\sigma_0)^2}$ \\  \bottomrule
\end{tabular}
\caption{{\bf Summary of thermoelectric coefficients } for the HLR and Dirac states. We have used units where $e=1$ and $\hbar=1$, so the quantum of resistance $h/e^2 = 2\pi$. See Eq.~(\ref{definition}) for the definition of $\sigma_0,\,\alpha_0$. 
\label{t:summary}}
\end{table*}

Eliminating the emergent electric field, $\v{e}$ and $\v{j}_\text{cf}$, from Eq.~(\ref{eq:je}) using Eqs.~(\ref{eq:jD}),(\ref{eq:cfresponse}), we find: $\v{j}_{el} = \[\frac{1}{4\pi}\hat\eps +\frac{\rho_0}{\(4\pi\)^2}\]\v{E}-\[\frac{1}{4\pi}\rho_0\alpha_0\hat\eps  \nabla T\]$, from which we can identify the electronic conductivity and thermoelectric susceptibility for the composite Dirac liquid:
\begin{align}
\hat\sigma &= \frac{\rho_0}{\(4\pi\)^2}\hat{1}
+\frac{1}{4\pi}\hat\eps
\hspace{0.25in}
\hat\alpha = \frac{\rho_0\alpha_0}{4\pi}\hat\eps
\label{eq:PHSresponse}
\end{align}
where $\rho_0 = \sigma_0^{-1}$ is the resistivity of the composite Fermi surface, $i,j\in\{x,y\}$.

Rather than observing $\hat\alpha$ directly, one typically measures the electric field, $\v{E}=\hat S\nabla T$, induced by a thermal gradient in a geometry where no electric current is permitted to flow: $\v{j}_{el} = \hat\sigma \v{E}-\hat\alpha\nabla T=0$, which defines the Seebeck tensor:
\begin{align}
\hat S &= \hat\sigma^{-1}\hat\alpha = \frac{\rho_0\alpha_0}{1+\(\rho_0/4\pi\)^2}\(\hat 1+\frac{\rho_0}{4\pi}\hat\eps\)
\end{align}
For a clean system ($k_F\ell \gg 1$), ${\rho_\text{0}}/({4\pi})\ll 1$, and the diagonal Seebeck coefficient $S^{xx}\approx \rho_0\alpha_0$, is essentially equal to the effective Seebeck coefficient that would be obtained by viewing the composite fermions as electrons, in reasonable quantitative agreement with previous analysis of thermopower measurements in GaAs~\cite{Ying94,Bayot95,Cooper97}. Note that this large Seebeck coefficient arises despite the particle-hole symmetry of the composite Dirac liquid. 

Moreover, we see that the off-diagonal component, $S^{xy}\approx {S^{xx}}/({k_F\ell})$ and closely related Nernst coefficient $\nu_N = S^{xy}/B$ are non-zero. We will shortly see that the non-vanishing $S^{xy}$ marks a sharp departure from the maximally particle-hole asymmetric HLR state with zero Berry phase, for which $S^{xy}_\text{HLR}=0$. More generally, we will show that the non-vanishing Nernst effect is directly tied to the Berry phase of the composite Fermi surface.

\header{Thermopower with zero Berry phase}
Composite fermions in the PHS breaking state originally described by HLR~\cite{HLR} are spinless electrons bound to $-4\pi$ flux, and hence carry electrical charge (equivalently flux of $\v{a}$) as well as vorticity. Then, in contrast to the Dirac liquid, the CF current in the HLR state is equal to the electron current, $\v{j}_\text{cf} = \v{j}_{el} = \frac{1}{4\pi}\hat\eps\(\v{E}-\v{e}\)$. In a thermopower measurement (i.e. in an open circuit configuration), zero electrical current flows, $\v{j}_\text{el}=0$, implying $\v{E}=\v{e}$, and $\v{j}_\text{cf}=0=\hat\sigma_\text{cf}\v{e}-\hat\alpha_\text{cf}\nabla T$, which together give $\v{E} = \hat\rho_\text{cf}\hat\alpha_\text{cf}\nabla T$, i.e. the electron and CF thermopower tensors coincide: $\hat{S}_{\text{HLR}} = \hat\rho_\text{cf}\hat\alpha_\text{cf}$. Since the CFs in the HLR state experience no residual orbital magnetic field, and if the composite Fermi-surface has zero Berry curvature, then the low-energy composite Fermi surface dynamics are effectively time-reversal symmetric such that $\hat\rho_\text{cf}$, $\hat\alpha_\text{cf}$ and hence also $\hat S$ to be purely diagonal (see Table~\ref{t:summary} for detailed comparison between HLR and Dirac).

We note that inelastic scattering from gauge fluctuations, whose propagator contains a Chern-Simons term and hence does not respect this fine-tuned spinless time-reversal, and can produce an off-diagonal component to $\sigma_{cf}$ and corresponding non-zero Nernst signal. However, neglecting extrinsic impurity effects, the transport scattering rate due to inelastic gauge scattering rate, $\Gamma_{g,\text{tr}}$, is expected to be suppressed at low temperatures by powers of $T$ (e.g. previous estimates~\cite{Lee92,HLR} give $\Gamma_{g,\text{tr}}\sim T^{2}$ for unscreened Coloumb interactions and $\Gamma_{g,\text{tr}}\sim T^{4/3}$ with interactions screened by a nearby gate). Then, the resulting gauge-fluctuation induced thermopower $S_g^{xy}\sim T \Gamma_{g,\text{tr}}$ can be distinguished by its temperature dependence from that arising from Berry phase in the composite Dirac liquid, which depends linearly on $T$. Thus, the absence of Berry curvature in the HLR state directly leads to a vanishing ($T$-linear component of) the Nernst signal.

\header{Effect of PHS breaking}
In experimental systems, inter-LL mixing is always present to some degree due to Coulomb interactions. However, at large fields for which the inter-LL spacing $\omega_c\approx \frac{eB}{m^*c}$ is much larger than the interaction energy $E_C\approx \frac{e^2 k_F}{\eps_r}$, one expects only weak inter-LL mixing and hence an approximate PHS. In this case, the composite Fermi-surface Berry phase is not quantized precisely to $\pi$, but should still deviate from $\pi$ only by a small amount $\approx \frac{E_C}{\omega_c}\ll 1$. Given this approximate PHS, it is natural to expect that the composite Fermi surface in 1/2-filled GaAs is approximately described by the composite Dirac theory, and may have appreciable Berry phase near $\pi$. 

To compare the thermoelectric responses of the particle-hole symmetric composite Dirac liquid, obtained above, to those of a particle-hole asymmetric state, we can include a particle-hole symmetry breaking Dirac mass ($m\neq 0$) in Eq.~(\ref{eq:Leff}), which produces a half-integer Hall conductance for the gauge field $a$. Physically, this PH breaking mass arises from inter-LL mixing, and we expect $m\approx \frac{E_C}{\omega_c}$.  

For non-zero PHS breaking mass, $m\neq 0$, the completely filled valence band of the composite Dirac cone contributes 1/2-integer quantized Hall conductance $\sigma_{D,v}^{xy}=\frac{\text{sgn}(m)}{4\pi}$, and the non-trivial Berry curvature enclosed by the conduction Fermi surface (FS) gives anomalous CF Hall conductance, $\sigma^{xy}_{D,\text{FS}}=\int_{k<k_F}\frac{d^2k}{(2\pi)^2}\Omega_k  = -\frac{\text{sgn}(m)}{4\pi}\gamma$, where $\Omega_k = \frac{m}{2\[(vk)^2+m^2\]^{3/2}}$ is the conduction band Berry curvature, and $\gamma \equiv \theta_B/\pi=1-\frac{m}{\sqrt{(vk_F)^2+m^2}}$ is a convenient measure of the composite FS Berry phase, $\theta_B$, which interpolates between $0$ (HLR) and $1$ (Dirac). The non-zero Berry curvature also induces anomalous CF thermoelectric Hall conductivity~\cite{Takehito11}, generically related to the CF number conductivity by the Mott formula: $\alpha_\text{D}^{ij} \approx \frac{\footnotesize\L T}{v_F} \frac{\d\sigma^{ij}}{\d k_F}$\cite{AHERMP,Xiao06,Qin11}.

Hence, for non-zero PHS breaking mass, $m$, $\sigma_D$ in Eq.~(\ref{eq:cfresponse}) becomes:
\begin{align}
\hat\sigma_{D}(m\neq 0) = \sigma_0\hat 1 +\text{sgn}(m)\frac{(1-\gamma)}{4\pi}\hat\eps
\end{align}
Then, proceeding as before by eliminating the internal electric field $\v{e}$ using Eqs.~(\ref{eq:jD}),(\ref{eq:cfresponse}), we find the electron conductivity and thermopower tensors for $m\neq 0$:
\begin{align}
\sigma &= \[\(\hat\sigma_\text{D}-\frac{\hat\eps}{4\pi}\)^{-1}-4\pi\hat\eps\]^{-1}\nonumber\\
S &= \(\hat\sigma_\text{D}-\frac{\hat\eps}{4\pi}\)^{-1}\hat\alpha_\text{D}
\label{eq:PHbreakingresponse}
\end{align}
The original HLR theory is recovered in the limit of zero Berry phase ($m\rightarrow \infty$). As previously remarked, in this limit the CF sector has a fine-tuned ``spinless" time-reversal symmetry, such that $\sigma_\text{cf}^{xy}=0$ and $\alpha_\text{cf}^{xy}\approx \frac{\scriptsize{\L} T}{v_F}\frac{\d\sigma_\text{cf}^{xy}}{\d k_F}=0$, resulting in vanishing Nernst coefficient.

For clean systems, $k_F\ell\gg 1$, $\sigma_0\gg \frac{1}{4\pi}$, the Hall conductivity and Nernst coefficient for generic composite Fermi surface Berry phase at half-filling are:
\begin{align}
\sigma^{xy} &\approx \frac{1}{4\pi}\(1-\text{sgn}(m)\frac{1-\gamma}{\(4\pi\sigma_0\)^2}\)
\nonumber\\
S^{xy} &\approx \L T\frac{k_F}{4\pi \sigma_0 v_F}\[\frac{\d\ln\sigma_0}{\d\ln k_F}+\text{sgn}(m)\frac{\d \ln\gamma}{\d\ln k_F}\] \times \gamma
\label{eq:Nernst}
\end{align}
where, for the massive Dirac model of Eq.~(\ref{eq:Leff}), $\frac{\d\ln\gamma}{\d \ln k_F} = (1-\gamma)(2-\gamma)$. Therefore, we see that the Nernst signal is directly sensitive to the Berry phase $\theta_B = \pi \gamma$.  Note that the Hall conductance is precisely ${e^2}/{(2h)}$ in the PHS theory with $\gamma=1$ ($\theta_B=\pi$), and otherwise deviates from this quantized value. In principle, this deviation can be used to distinguish the Dirac and PH breaking CFL states~\cite{Son15}, however in practice this deviation is a very small correction, of order $\mathcal{O}\(k_F\ell\)^{-2}$, to the total Hall conductance, and will be challenging to measure. In contrast, we see that the Nernst signal is directly proportional to the composite Fermi surface Berry phase, and is thus a direct measurement (rather than small correction) of the topological properties of the composite Fermi surface.

To extract $\theta_B$ from a Nernst effect measurement, it is desirable to independently determine the relevant parameters in Eq.~(\ref{eq:Nernst}) through other transport measurements, rather than relying on theoretical calculation in some particular model. To this end, $\sigma_0$ can be obtained by the longitudinal electrical conductivity $\sigma^{xx}\approx \frac{1}{(4\pi)^2\sigma_0}$, and the logarithmic derivative of $\sigma_0$ can be obtained from the longitudinal thermopower measurement $S^{xx}\approx \frac{\scriptsize{\L}Tk_F}{v_F}\frac{\d \ln \sigma_0}{\d \ln k_F}$. Together, these measurements allow for an experimental determination of the composite Fermi-surface Berry phase via the expression (valid when $\theta_B$ is close to $\pi$):
\begin{align}
\theta_B \approx \frac{S^{xy}}{4\sigma^{xx}S^{xx}}
\end{align}
in a manner that is largely insensitive to the details of non-universal features such as composite Fermi velocity and impurity scattering rates. Here, we remind that all conductivities are written in units ${e^2}/{\hbar}$.

\begin{figure}[tb]
\includegraphics[width=0.7\columnwidth]{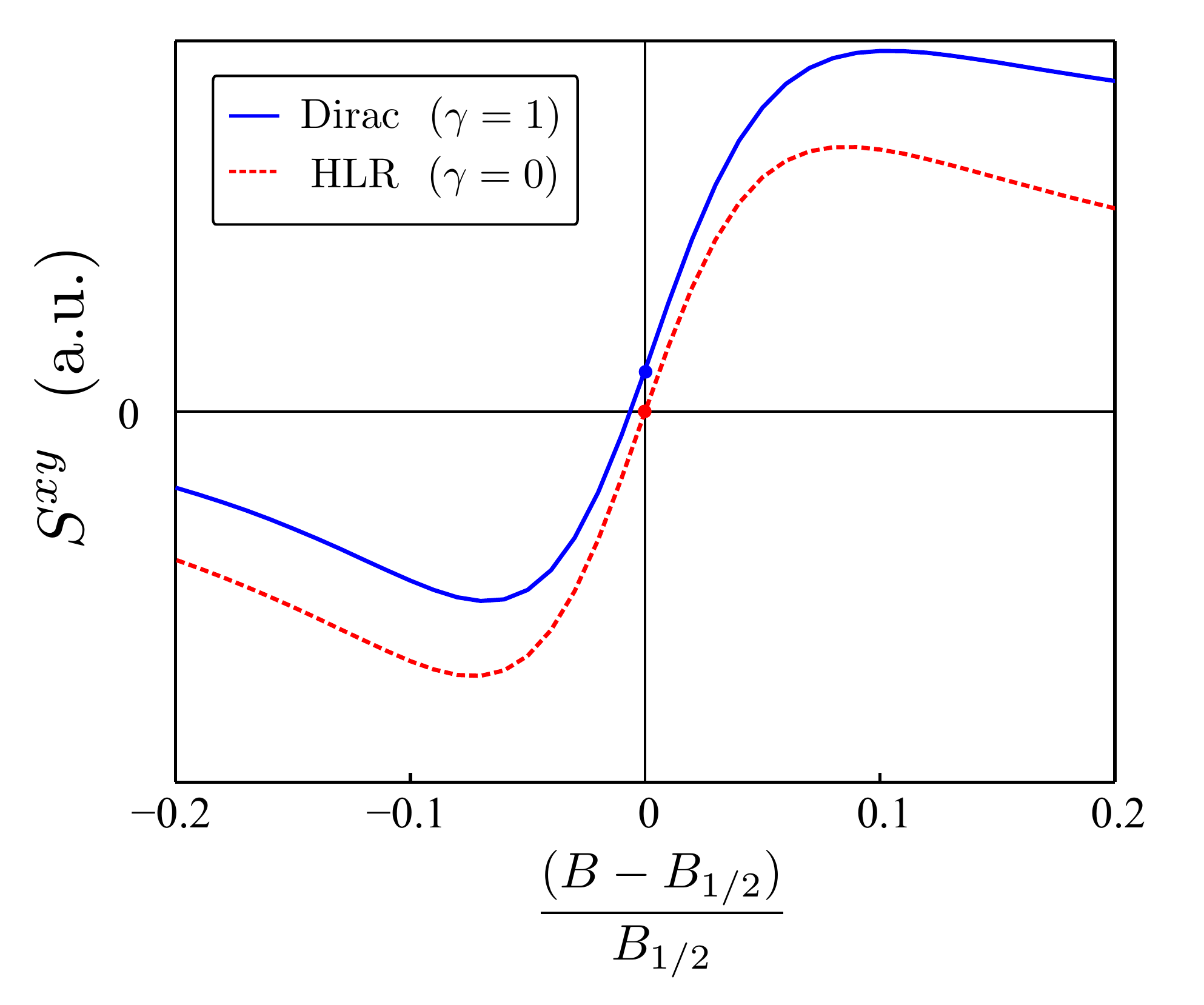}\vspace{-.1in}
\caption{\label{fig:NernstPlot} {\bf Field dependence of $S^{xy}$ - } In addition to whether $S^{xy}$ vanishes at half-filling ($B=B_{1/2}$), the asymmetric field dependence of the transverse thermopower (Nernst) signal, distinguishes between the HLR state with zero Berry phase ($\gamma=0$), and the particle-hole symmetric Dirac state with $\pi$ Berry phase ($\gamma=1$).}
\label{fig:NernstField}
\vspace{-.1in}
\end{figure}

\header{Field detuning}
We have shown that the Nernst coefficient at half-filling provides a sensitive probe for the composite Fermi surface Berry phase. In addition, the Dirac and HLR states also show distinct Nernst signals upon tuning the field away from half-filling, $B\neq B_{1/2}\equiv 4\pi n_{el}$, where $n_{el}$ is the electron density. For $B\neq B_{1/2}$, an effective gauge-magnetic field $b = B-B_{1/2}$ arises, which induces a change in the CF Hall conductance. For non-zero composite FS Berry phase $\theta_B=\pi\gamma$, the field-detuning alters the CF density by: $n_\text{cf} = \frac{B-(1-\gamma)b}{4\pi}$, and the composite Fermi wavevector is $k_F = \sqrt{4\pi n_\text{cf}}$. Let us work in the limit of  small detuning, $b\ll B_{1/2}$, where many Landau levels of the composite Fermi sea are occupied, and the system is still compressible due to disorder broadening of the CF LLs. In this regime, the CF conductivity becomes:
\begin{align}
\hat\sigma_\text{D} = \sigma_0\frac{\(\hat1+\omega_c\tau\hat\eps\)}{1+(\omega_c\tau)^2}+\frac{1-\gamma}{4\pi}\hat\eps
\label{eq:SigmaField}
\end{align}
where $\omega_c = \frac{v_Fb/k_F}{1+b\Omega_F}$ is the CF cyclotron frequency, $\Omega_F$ is the Berry curvature at the composite Fermi surface. Since the transport distinctions between Dirac and PH-breaking states arise from the difference between $\gamma=0,1$, from Eq.~\ref{eq:SigmaField}, we see that this distinction is lost unless: $\frac{|B-B_{1/2}|}{B_{1/2}} < \(k_F\ell\)^{-2}$. Thus accurately determining $B_{1/2}$ is an important component of any transport measurement designed to observe signatures of CF Berry phase. In the impurity scattering dominated (non-hydrodnamic) regime $\alpha_\text{cf}\approx \L T\frac{\d \sigma_\text{cf}}{\d \e_F}$. To estimate this quantity, we need to assume a particular energy dependence for the scattering rate $\tau\sim \e_F^p$, e.g. for free electrons with short-range impurities $p=1$, however, the precise value of $p$ does not qualitatively affect the results. Then, we may compute the field dependence of the Nernst coefficient using Eqs.~(\ref{eq:PHbreakingresponse}),(\ref{eq:SigmaField}). The results are shown in Fig.~\ref{fig:NernstField} with $k_F\ell=50$ for zero and $\pi$ Berry phase (HLR, Dirac respectively). Both curves increase in magnitude towards a maximum at $\omega_c\tau\sim 1$, beyond which, the Nernst coefficient decays as $\sim \frac{1}{\omega_c\tau}$. In addition to the non-zero value of $S^{xy}$ at $B=B_{1/2}$
the Dirac CFL shows a distinct asymmetry between $B>B_{1/2}$ and $B<B_{1/2}$ due to the intrinsic anomalous ($b=0$) Hall contributions from $\gamma\neq 0$. 

\header{Experimental considerations}
Based on data from~\cite{Ying94}, we estimate the size of the non-zero $S^{xy}$ signal at half-filling in the composite Dirac liquid will be of order $0.3~\mu$V/K at $T=100$~mK. This magnitude, while somewhat small, should be measurable in existing thermopower setups.  An additional experimental challenge is that the longitudinal and transverse thermopower signals may be mixed to a small degree due to non-idealities in the measurement geometry, and need to be disentangled. The two components can be separated by noting that the transverse response, $S^{xy}$, will be odd under $B\rightarrow -B$, whereas the longitudinal component will be even. Hence, antisymmetrizing (or symmetrizing) the thermopower signal under  $B\rightarrow -B$ can help isolate $S^{xy}$ (or $S^{xx}$) respectively.

While the smallness of the Nernst signal may make the proposed thermopower signature more challenging to measure compared to previously proposed~\cite{Son15} deviation of electrical Hall conductance from $1/2$, the thermopower setup has a distinct advantage. Namely, whereas $\sigma^{xy}=1/2$ would follow directly from the constraint of particle-hole symmetry, and does not distinguish the Dirac composite Fermi liquid theory from other putative particle-hole symmetric states, the non-zero linear-T component of the Nernst signal is not a direct consequence of symmetry, but rather a specific prediction for the composite Dirac liquid state. Hence its observation is more directly tied to the existence of a composite Fermi surface with non-zero Berry phase. In practice, the strongest experimental evidence for Berry phase structure of the composite Fermi surface would come from looking for consistency with a number of different electrical and thermoelectric transport predictions.

\header{Discussion}
To summarize, we have analyzed the thermopower of composite Fermi liquid (CFL) states with and without particle hole symmetry, and demonstrated that the Nernst effect is directly sensitive to the Berry curvature of the composite Fermi surface, and can be used to distinguish between previously HLR and Dirac theories of the CFL. The computations of this paper amount to a self-consistent RPA treatment of gauge fluctuations, in the spirit of Ioffe and Larkin's method for computing the transport response of fractionalized phases~\cite{IoffeLarkin}. This approach is formally justified in the limit of a large number of fermion flavors, however, since the effects we predict are largely a consequence of symmetry considerations, we expect our results to be more generally valid. Additionally, we have evaluated the response functions in the limit where elastic impurity scattering dominates over inelastic gauge-field mediated interactions. Potentially interesting hydrodynamic effects in the inelastic dominated regime require more sophisticated treatments, and are left to future work.

\vspace{4pt}\noindent{\it Acknowledgements -- } We thank B.~I. Halperin, N. Cooper, C. Wang, J. Alicea and M. Zaletel for insightful conversations. ACP and MS were supported by the Gordon and Betty Moore Foundation's EPiQS Initiative through Grant GBMF4307, AV was supported by a Simons Investigator grant.

\appendix

\section{Parton description of composite Fermi liquid(s) \label{app:Partons}}
It is convenient to access the low-energy field theory of the composite Fermi liquid via a parton (``slave-particle") construction that is capable of describing both the PHS composite Dirac- and PH breaking HLR state on similar footing. Namely, consider decomposing the electron into a charge-1 boson, $b$, and a neutral fermion, $f$: $c=bf$, and constraining $b$ and $f$ to appear together in physical correlation functions by introducing a $U(1)$ gauge field $a_\mu$ under which $b$ and $f$ have charge $\pm 1$ respectively. This corresponds to an effective field theory with imaginary time Lagrangian density: $\mathcal{L}=\mathcal{L}_b+\mathcal{L}_f-(A+a)_\mu j^\mu_b-(-a_\mu) j^\mu_f$, where $\mathcal{L}_{b,f}$ are the Lagrangian densities and $j_{b,f}$ are the number currents of $b$ and $f$ respectively. 

Composite Fermi liquid states of the 1/2-filled Landau level can be obtained by having the bosonic partons, which are at $1/2$-filling with respect to the external magnetic flux, form an incompressible abelian (Laughlin-like) fractional quantum Hall state, $\mathcal{L}_b = -\frac{2i}{4\pi}a_bda_b$, where $a_b$ is a vector field whose flux is the boson current: $j_b = \frac{da_b}{2\pi}$. This mean-field ansatz is equivalent to the ``flux-attachment" picture typically utilized in composite fermion descriptions of quantum Hall states, as each boson in the $\nu=1/2$ FQH state is accompanied by $4\pi$ flux of $a$ due to their quantum Hall response, which the composite fermions $f = b^\dagger c$ see has $-4\pi$ flux. The fermionic partons, $f$, then see zero average magnetic flux, and may form a simple Fermi-surface (this is the usual HLR theory, which is maximally particle-hole symmetry violating). Alternatively, we may access the PHS composite Dirac state by putting $f$ into a band-structure consisting of two Dirac cones, with flavor index $i=1,2$: 
\begin{align}
\mathcal{L}_f = \sum_{i=1,2}\psi^\dagger_{a,i}\(i\d_\tau-\mu+v\v{p}\times \bs{\sigma}_{ab}-m_i\sigma^z_{ab}\)\psi_{b,i}
\label{eq:Lf}
\end{align}
a PHS low-energy theory is obtained by keeping one flavor of massless Dirac fermions $m_1=0$, and giving the other a large, negative mass $(m_2<-\mu<0)$. More generally, we may interpolate between the two theories by continually varying $m_1=m$ between $0$ (Dirac) and $\infty$ (HLR), subject to $0<m<\mu<|m_2|$.

We caution that Eq.~(\ref{eq:Lf}) can only hope to capture the structure of low-energy excitations close to the composite Fermi surface of $\psi_{a,1}$, since excitations at this energies will involve non-universal short-distance physics outside the purview of the effective field theory of the composite (marginal) Fermi-liquid state. Hence, one should not consider the detailed high energy structure, such as linearity or curvature of the CF dispersion far from $\mu$, to be physically meaningful.

\section{Linear response from parton theory \label{app:PartonResponse}}
We now turn to a computation of the (linear) electric current response to (small) applied voltages and thermal gradients. 

\subsection{Preliminaries}
The proper computation of such transport coefficients faces several subtleties. Chiefly, one must properly distinguish between ``free" or ``transport" currents that can pass freely between conducting leads and the sample, and circulating (i.e. have zero divergence) magnetization currents that are bound to the sample. While Kubo formulae typically yield expressions for the total (transport plus magnetization) current, in a transport measurement, only the former are observable and the magnetization contribution must be properly subtracted in calculations. 

Moreover, electric currents can be generated either by applying electric fields that exert mechanical forces to accelerate electrons or by setting up density gradients that exert ``statistical" forces on electrons, e.g. leading to diffusion. In particular, for slowly varying electric potentials and densities that are close to equilibrium values, the local density matrix near position $\v{r}$ (coarse grained over a distance much larger than Fermi-wavelength, but much smaller than the characteristic lengthscale for potential gradients) is well describe by a local equilibrium form: $\rho(\v{r})\approx \frac{1}{\mathcal{Z}(r)}\e^{-\beta(\v{r})\(\hat{H}+\tilde{\mu}(\v{r})n(\v{r})\)}$. Here, the local ``chemical potential" $\tilde{\mu}(\v{r})=\mu(\v{r})-\phi(\v{r}) = -\frac{\d\Omega}{d n(\v{r})}$ is the difference between the electrochemical potential, $\mu$ (which is uniform in equilibrium), and the electrostatic potential $\phi(\v{r})$ (related to electric field by $\v{E}=-\nabla\phi$), here $\Omega=-T\log\mathcal{Z}$ is the free-energy. Since the partition function in the presence of $\nabla\tilde{\mu}$ can be related to that with a electric field $\nabla\phi=\nabla\tilde\mu$ by a time-dependent gauge transformation $c(\v{r},t)\rightarrow e^{-i\tilde{\mu}(\v{r})t}c(\v{r},t)$, the transport current generated by $\tilde{\mu}$ must be equal to that generated by $\nabla\phi$. This leads to the familiar Einstein relation for transport currents, which requires that $\tilde{\mu},\phi$ enter into the transport current via the combination $\nabla\mu = \nabla\tilde\mu-\nabla\phi$. Similarly, one may drive currents via thermoelectric response either by a temperature gradient $\nabla T$, the analog of the statistical force $\nabla\tilde\mu$, or by coupling the system to a ``gravitational" potential $\psi(\v{r})$, coupled to the local Hamiltonian density~\cite{Luttinger64} whose gradient mechanically accelerates electrons (analogous to $\phi$). Similar to the Einstein relation for $\tilde{\mu},\phi$, the transport current depends only on the combination $\nabla\psi-\frac{1}{T}\nabla T$.

While in a typical experiment, one applies voltage, $-\nabla\mu$, and temperature gradient $\nabla T$, theoretical computations with these mixed mechanical and statistical forces are complicated by the presence of $\tilde{\mu}$ and $T$ gradients that give rise to magnetization currents $J^M=\frac{\d M}{\d \tilde{\mu}}\times\nabla\tilde{\mu}+\frac{\d M}{\d \tilde{T}}\times\nabla\tilde{T}$, which must be carefully subtracted to obtain the transport current. It is instead simpler to follow the approach of Cooper, Halperin, and Ruzin~\cite{Cooper97}, and compute the linear response current for mechanical potentials $\nabla\phi,\nabla\psi$ with uniform $\tilde{\mu},T$ (e.g. using Kubo formulae), and then later restore $\nabla\tilde{\mu},\nabla T$ using Einstein relations. This approach has the virtue that the change in magnetization for $\nabla\tilde\mu=0=\nabla T$, $\delta m(\v{r})=-\frac{\d\Omega}{\d B(\v{r})}=\psi(\v{r})m(\v{r})$, arises only from the re-weighting of the local energy density by $\(1+\psi(\v{r})\)$. The corresponding magnetization current: $\v{J}^M=m(\v{r})\hat\eps\(-\nabla\psi\)$, can then be readily subtracted from the total current to yield the transport current $\v{j}^\text{tr}(\nabla\phi,\nabla\psi) = \[\v{J}^\text{tot}-\v{J}^M\](\nabla\phi,\nabla\psi)$. The transport current for generic mechanical and statistical potentials can then be obtained by utilizing Einstein relations, to substitute $\phi\rightarrow \phi-\mu$, and $\nabla\psi\rightarrow \nabla\psi-\frac{1}{T}\nabla T$. This procedure automatically ensures that the resulting transport current satisfy Einstein relations and Onsager reciprocity conditions, and simplifies the subtraction of magnetization currents.

\subsection{General Ioffe-Larkin composition rules for thermoelectric transport coefficients}
Then, quite generally, the thermoelectric response theory of the state described by any parton ansatz (i.e. specific choice of $\mathcal{L}_{b,f}$) can be obtained by coupling the system to a gravitational potential $\nabla \psi$ and external field with vector potential $A_\text{ext} = A_B+A$ (here $\nabla\times \v{A}_B=\v{B}$ is the vector potential corresponding to the applied magnetic field to produce 1/2-filling, and $A$ is a small additional probe gauge field), and integrating out the matter fields $b$ and $f$ to obtain the linear response action:
\begin{widetext}
\begin{align}
\mathcal{L}_\text{LR}[A,a] =& -(A+a)_\mu\[\frac{1}{2}K^{\mu\nu}_b(A+a)_\nu + \tilde\alpha_b^{\mu\nu}\(-T\d_\nu \psi\)+\(J_b^M\)_\nu\]- a_\mu\[\frac{1}{2}K^{\mu\nu}_fa_\nu+\tilde\alpha_f^{\mu\nu}\(-T\d_\nu \psi\)-\(J_f^M\)_\nu\]
\end{align}
\end{widetext}
where $K_{b,f}$ ($\tilde\alpha_{b,f}$) are the gauge (thermoelectric) response kernels of conserved $b$ and $f$ number currents respectively. For example, the number conductivity of $b,f$ is $\sigma^{ij}_{b,f}(\omega,\v{q}) = \frac{1}{i\omega}K^{ij}_{b,f}(\omega,\v{q})$. $\(J_{b,f}^{M}\)^i =  \eps^{ij}\d_j M_{b,f}$ are the $b$ and $f$ number magnetization currents respectively, and similarly $M_{b,f}$ are the $b,f$ number magnetizations respectively. These quantities can be computed for any specific parton ansatz, and we will give explicit expressions for for composite Fermi liquid states below. However, for the moment we leave $K,\alpha,J^M$ unspecified to derive general formulae.

Next, we may enforce the gauge constraint by integrating out the constraint gauge field $a$ at the RPA level. This produces a linear response action for the physical electromagnetic field, $A$: $\mathcal{L}_\text{LR}[A] = -\frac{1}{2}AKA-A\tilde\alpha\(T\nabla\psi\)$, where we can identify the electronic conductivity, and thermoelectric response tensors:
\begin{align}
K &= \(K_b^{-1}+K_f^{-1}\) \nonumber\\
\tilde\alpha &= K\(K_b^{-1}\tilde\alpha_b+K_f^{-1}\tilde\alpha_f\) \nonumber\\
J^M & = K\(K_b^{-1}J^M_b+K_f^{-1}J^M_f\) 
\end{align}
The first expression is a straightforward generalization of the familiar Ioffe-Larkin composition rule for the electronic electromagnetic response kernel in terms of the parton response kernels~\cite{IoffeLarkin}. The latter two generically express the physical thermoelectric response, and magnetization current in terms of the corresponding parton responses.

Following the procedure outlined in the previous section, the magnetization induced by a gravitational potential (with $\nabla\tilde{\mu},\nabla T=0$) as: $J^M_{b,f}=M_{b,f}\hat\eps\nabla\psi$, so that the electronic transport current induced by $\nabla\psi$ is $j^\text{tr} = K\[K_b^{-1}\(\tilde\alpha_b-\frac{1}{T}M_b\hat\eps\)+K_b^{-1}\(\tilde\alpha_f-\frac{1}{T}M_f\hat\eps\)\]\(T\nabla\psi\)$, from which we identify the electronic thermoelectric tensor:
\begin{align}
\alpha = K\[K_b^{-1}\(\tilde\alpha_b-\frac{1}{T}M_b\hat\eps\)+K_f^{-1}\(\tilde\alpha_f-\frac{1}{T}M_f\hat\eps\)\]
\end{align}
We note, that the effect of properly subtracting the total electronic magnetization current, is identical to simply subtracting the parton number magnetization currents from the respective parton response coefficients: $\tilde{\alpha}_{b,f}\rightarrow \alpha_{b,f}=\tilde{\alpha}_{b,f}-\frac{M_{b,f}}{T} \hat\eps$. 

\subsection{Transport coefficients for the half-filled Landau level}
For both the Dirac and ordinary composite Fermi liquid theories, the bosonic partons can be taken to form an incompressible FQH state. Since $b$ forms a gapped FQH state, there is an energy gap, $\Delta$, for adding an electron (note that, while the CFL is compressible, this arises from gapless $f$ excitations that have zero overlap with electrons). For temperatures well below $\Delta$, we may neglect deviations from this ideal quantized Hall behavior described by: $K_b^{\mu\nu} = \frac{i}{4\pi}\eps^{\mu\lambda\nu}\d_\lambda$, and $\alpha_b = 0$. In particular, the boson number conductivity is quantized and purely off-diagonal: $\sigma^{ij}_b = \frac{K^{ij}(\omega\rightarrow 0,q=0)}{i\omega} = \frac{1}{4\pi}\eps^{ij}$

For the fermionic sector, we use the Dirac ansatz, (\ref{eq:Lf}), and allow a PH breaking mass $m_1\equiv m$ to tune between the composite Dirac and HLR states. Let us denote the CF number and thermoelectric conductivity of the compressible CF Fermi surface of fermions $\psi_{a,1}$ by $\sigma_{f,1}$ and $\alpha_{f,1}$ respectively, and similarly define $\sigma_{f,2}$, $\alpha_{f,2}$ for the incompressible ``regulator" fermions $\psi_{a,2}$. The total CF number conductivity is given by the sum of the number conductivities for each CF ``flavor":
\begin{align}
\hat\sigma_f = \sum_{i=1,2}\hat\sigma_{f,i}
\end{align}
The $\psi_{f,2}$ contribute a CF Hall conductance $\sigma_{f,2}=-\frac{1}{4\pi}\hat\eps$, which cancels the $ada$ type Chern-Simons term for the internal gauge field from the boson sector. However, since $\psi_{f,2}$ are incompressible they do not respond to thermal gradients and have $\alpha_{f,2}=0$, hence:
\begin{align}
\hat\alpha_f = \hat\alpha_{f,1}
\end{align}

We now turn to the contributions, $\hat\sigma_{f,1}$, $\hat\alpha_{f,1}$ from the compressible composite Fermi surface of $\psi_{a,1}$ fermions. At low temperatures, and long-wavelengths elastic impurity scattering dominates over inelastic scattering due to interactions (e.g. mediated by fluctuations of the internal gauge field $a$), and response coefficients for the $f$-sector are approximately given by those of a Fermi-liquid (albeit with renormalized effective parameters, $v_F, m,$ etc...). 

The transport coefficients for this state can computed, e.g. by using semiclassical equations of motions and Boltzmann type kinetic equations\cite{Xiao06,Qin11}:
\begin{align}
\sigma_{f,1}^{xx} &= \frac{k_F\ell}{4\pi} \hspace{0.3in}
\sigma_{f,1}^{xy} = \int\frac{d^2k}{(2\pi)^2}n_F(\e_k)\Omega_k 
\end{align}
where $\ell = v_F\tau$ is the mean-free path from scattering off impurities elastically with rate $\tau^{-1}$. Here, $\Omega_{\v{k}} = \eps^{ij}\d_{k^i}\<u_k|i\d_{k^j}|u_k\> = \frac{m}{2\e_k^{3/2}}$ is the Berry curvature density for the composite Fermi sea with dispersion $\e_{\v{k}} = \sqrt{(vk)^2+m^2}$, and $n_F(\e_k)$ is the Fermi-occupation factor. The $f$-thermoelectric tensor (including magnetization subtraction) was shown to satisfy the Mott rule\cite{Xiao06,Qin11}:
\begin{align}
\alpha_f^{ij} = \alpha_{f,1}^{ij}= \L T\frac{\d\sigma^{ij}_{f,1}}{\d \e_F} = \frac{\L T}{v_F}\frac{\d\sigma^{ij}_{f,1}}{\d k_F} 
\end{align}

\section{Massive Dirac model}
In this Appendix, we derive some useful relations for the massive Dirac model in Eq.~(\ref{eq:Leff}), for the composite Fermi surface with general particle-hole breaking mass. We note that only the low-energy excitations near the composite Fermi surface have any physical meaning, since high energy excitations deep below the Fermi surface represent non-universal short-distance physics which are beyond the applicability of the effective low-energy long-wavelength theory. For this reason, it is desirable to re-write the parameters $v,m,\mu$ in Eq.~(\ref{eq:Leff}), in terms of measurable low-energy quantities including Fermi velocity, $v_F = \left.\frac{\d\e}{\d k}\right|_{k=k_F} = v\gamma$, and Fermi-surface Berry phase: $\theta_B = \pi\gamma = \pi\(1- \frac{m}{\sqrt{(vk_F)^2+m^2}}\)$, as well as the Fermi wavevector, $k_F$. The latter is not a low-energy property, but is related to the density of electrons at half-filling, $k_F = \sqrt{4\pi n_\text{el}}$, and is therefore accurately known. The unphysical quantities can be replaced by physically measurable ones, using the relationships: 
\begin{align}
v &= \frac{v_F}{\sqrt{\gamma(2-\gamma)}},\hspace{0.25in}
m = v_Fk_F\frac{(1-\gamma)}{\gamma(2-\gamma)}
\nonumber\\
\mu &= \sqrt{(vk_F)^2+m^2} = \frac{v_Fk_F}{\gamma(2-\gamma)}
\end{align}

One can also compute the Berry curvature as a function of CF momentum $k$ for this model. The single-particle spinor wave-function for a CF with momentum $k$, is $u_k = \frac{1}{\sqrt{(\e_k-m)^2+(vk)^2}}\begin{pmatrix}e^{-i\phi_k}\(\e_k-m\)\\ ivk \end{pmatrix}$, where $\e_k = \sqrt{(vk)^2+m^2}$, and $\phi_k = \tan^{-1}\frac{k_y}{k_x}$. This gives a Berry connection $\mathcal{A}^{(B)}(\v{k}) = i\<u_{\v{k}}|\d_{\v{k}}|u_{\v{k}}\>$, and Berry curvature:
\begin{align}
\Omega_k = \eps^{ij}\d_{k^i}\mathcal{A}_j(\v{k}) = \frac{m}{2\[(vk)^2+m^2\]^{3/2}} 
\end{align}

\section{Impurities and extrinsic anomalous Hall response \label{app:Disorder}}
So far, we have ignored extrinsic sources of anomalous Hall conductance for the CF response coefficients due to disorder scattering. Specifically, while disorder preserves PHS on average, it can have higher odd cumulants that statistically bias scattering in a left- or right-handed fashion thereby producing hall conductance\cite{AHERMP}. In this Appendix, we examine disorder contributions to $\sigma_f^{xy}$ for the PHS composite Dirac liquid, and argue that they are negligible under realistic conditions for GaAs samples. The crucial point is that, PH breaking effects such as skew-scattering that contribute to $\sigma_f^{xy}$, arise only from higher order odd cumulants of the impurity distribution. In GaAs, the main source of impurities are dopants, that are separated from the two dimensional electron gas (2DEG) by a distance $d\gg k_F$. In this case, the scattering from higher cumulants of the disorder potential is suppressed by factors of $\frac{1}{k_Fd}\ll 1$, and hence are expected to be small.

For dopant impurities located at positions $\{\v{r}_i\}$ in the plane, and separated a distance $d$ from the 2DEG conduction layer in the $z$-direction, the bare impurity potential is: $V_0(\v{r})=\sum_i\frac{e}{\e_d\sqrt{|\v{r}-\v{r}_i|^2+d^2}}$ where $\e_d$ is the dielectric constant of GaAs. This bare potential will be screened by the 2DEG, and induce a screening charge whose Fourier components are given, in the Thomas-Fermi approximation by:
\begin{align}
\delta\rho_\text{ind}(\v{q})\approx e^{-qd}\delta\rho_\text{imp}(\v{q})
\end{align}
where $\delta\rho_\text{imp}(\v{q}) = \sum_i e^{i\v{q}\cdot\v{r}_i}-\rho_0\delta(\v{q})$ and $\v{r}_i$ is the position of the $i^\text{th}$ impurity, and $\rho_0=\frac{k_F^2}{4\pi}$ is the density of dopants (also the density of electrons).

The composite Dirac fermions see inhomogeneities in this induced charge as a random magnetic field of strength $b_\text{imp}(\v{r})=4\pi\rho_\text{ind}(\v{r})$. This impurity field vanishes on average, but has non-trivial higher moments that scale like $\overline{b_\text{imp}^n} \approx \frac{\rho_0}{d^{2n}}$, which can contribute extrinsic sources of anomalous Hall conductance for the composite fermions. However, we will subsequently estimate that these extrinsic Hall contributions are suppressed by a factor of $\frac{1}{k_Fd}\ll 1$, and can be neglected.

A subtlety in this analysis, is that the effects of random orbital magnetic fields are inherently non-analytic in the field strength and cannot be treated by standard perturbative Green function expansions\cite{Altshuler92}. The mean free path can be estimated by considering a semiclassical treatment of the conductivity bubble in real space: $\sigma(r,r') = \lim_{\omega\rightarrow 0} \overline{\frac{1}{i\omega}\int dt e^{-i\omega t}\text{tr}\[v^i G^r(\v{r},\v{r}';t)v^i G^a(\v{r}',\v{r};0)\]}$, where $G^{r,a}$ are the retarded and advanced composite fermion Green functions respectively, and $v^i$ is the $i^\text{th}$ component of the velocity operator. The conductivity bubble may be viewed as the propagation of a particle from $\v{r}\rightarrow \v{r}'$ and then returning from $\v{r}'\rightarrow \v{r}$. For $k_F|\v{r}-\v{r}'|\gg 1$, the propagation is well approximated by a sum over paths weighted by gaussian amplitude in the deviation from the classical path (a straight line from $\v{r}\rightarrow \v{r}'$), and with phase given by the magnetic flux enclosed in the particle propagating from $\v{r}$ to $\v{r}'$ and back. In this semiclassical approximation, quantum fluctuations in the path are of typically RMS size $\sqrt{|\v{r}-\v{r}'|\lambda_F}$, where $\lambda_F = \frac{2\pi}{k_F}$ is the wavelength of the particle, corresponding to typical area: $\mathcal{A}\approx |\v{r}-\v{r}'|^{3/2}\sqrt{\lambda_F}$ i.e. flux $\Phi\approx b_\text{rms}\mathcal{A}$, where $b_\text{rms}\approx\sqrt{\overline{b_\text{imp}^2}}\approx k_F/d$ is the RMS size of the dual magnetic field experienced by the CFs due to impurities. When $|\v{r}-\v{r}'|$ is sufficiently large that $\Phi\approx 2\pi$, the sum over quantum paths contains a rapidly fluctuating phase. Beyond this length scale, the conductivity decays exponentially with separation between $|\v{r}-\v{r}'|$, from which one can identify the mean-free path $\ell\approx \(b_\text{rms}^2\lambda_f\)^{-1/3}\approx \(d^2/k_F\)^{1/3}$ (note that $k_F\ell\approx \(k_f d\)^{2/3}$ is naturally $\gg 1$ if the distance to the doping layer is much larger than the inter-electron spacing).

To estimate the skew scattering contributions to anomalous CF Hall conductivity, we can compute the statistical bias for an electron to deflect to its left rather than its right within a mean-free path, $\ell$. Consider a CF initially traveling at speed $v_F$ along the $x$-direction. Due to the random local impurity field, $b_\text{imp}(\v{r})$, the CF experiences a Lorentz force and accelerates in the y-direction. The dual field is roughly constant over distances of the mean-free path, $\ell$. During the scattering time $\tau\approx \frac{\ell}{v_F}$, the CF experiences cyclotron motion under the local field $b$, and develops a perpendicular momentum: $\delta k_\perp(\tau) \approx k_F \sin \omega_c\tau$ in the y-direction.
Though, the impurity averaged Lorentz force is zero, the non-linearity in the $\sin$ function in the cyclotron motion of the electron give a non-zero average to $\delta k_\perp$:
\begin{align}
\overline{\(\frac{\delta k_\perp(\tau)}{k_F}\)} \approx \frac{1}{3!}\overline{b^3}\(\frac{\ell}{k_F}\)^3\approx \frac{1}{(k_F d)^2}\ll 1
\end{align}
However, the amplitude of this skew-scattering bias is strongly suppressed by the smooth character of the impurity potential, and should be negligible in practice.

While the smooth nature of the impurity potential strongly suppresses skew-scattering events, it also may lead to departures from ordinary Fermi-liquid like behavior for the CFs. Namely, the typical CF cyclotron radius due to the dual magnetic field induced by impurities, $R_c = \frac{b_\text{rms}}{k_F}\approx d$, is comparable to the typical length scale, $d$, of spatial variations in $b_\text{imp}(\v{r})$. Whereas for $\frac{R_c}{d}\gg 1$, the CF trajectory is only weakly affected by the impurity field, in the opposite limit, $\frac{R_c}{d}\ll 1$, the CFs motion is better described by rapid cyclotron orbits with slowly drifting guiding center\cite{Mirlin98}. In the latter regime, conduction occurs by percolating ``snake" state channels along contours of $b_\text{imp}(\v{r})=0$, and is qualitatively different than for an ordinary metal with a sharp Fermi surface. The above described impurity model suggests that typical GaAs samples lie in an intermediate regime between these extreme limits. It is possible to avoid this potential issues by using compensated dopants consisting of both donors and acceptors of density $n_{d,a}$ respectively. Then, if the conduction electron density $n_{el} = n_d-n_a$, is much less than the total number of dopants $n_\text{imp} = n_d+n_a$, $R_c/d\approx \sqrt{n_{el}/n_\text{imp}}$ can be made much smaller than one, ensuring that long-wavelength impurity $b$-field fluctuations do not alter the usual composite Fermi surface picture.


\begin{thebibliography}{26}%
\makeatletter
\providecommand \@ifxundefined [1]{%
 \@ifx{#1\undefined}
}%
\providecommand \@ifnum [1]{%
 \ifnum #1\expandafter \@firstoftwo
 \else \expandafter \@secondoftwo
 \fi
}%
\providecommand \@ifx [1]{%
 \ifx #1\expandafter \@firstoftwo
 \else \expandafter \@secondoftwo
 \fi
}%
\providecommand \natexlab [1]{#1}%
\providecommand \enquote  [1]{``#1''}%
\providecommand \bibnamefont  [1]{#1}%
\providecommand \bibfnamefont [1]{#1}%
\providecommand \citenamefont [1]{#1}%
\providecommand \href@noop [0]{\@secondoftwo}%
\providecommand \href [0]{\begingroup \@sanitize@url \@href}%
\providecommand \@href[1]{\@@startlink{#1}\@@href}%
\providecommand \@@href[1]{\endgroup#1\@@endlink}%
\providecommand \@sanitize@url [0]{\catcode `\\12\catcode `\$12\catcode
  `\&12\catcode `\#12\catcode `\^12\catcode `\_12\catcode `\%12\relax}%
\providecommand \@@startlink[1]{}%
\providecommand \@@endlink[0]{}%
\providecommand \url  [0]{\begingroup\@sanitize@url \@url }%
\providecommand \@url [1]{\endgroup\@href {#1}{\urlprefix }}%
\providecommand \urlprefix  [0]{URL }%
\providecommand \Eprint [0]{\href }%
\providecommand \doibase [0]{http://dx.doi.org/}%
\providecommand \selectlanguage [0]{\@gobble}%
\providecommand \bibinfo  [0]{\@secondoftwo}%
\providecommand \bibfield  [0]{\@secondoftwo}%
\providecommand \translation [1]{[#1]}%
\providecommand \BibitemOpen [0]{}%
\providecommand \bibitemStop [0]{}%
\providecommand \bibitemNoStop [0]{.\EOS\space}%
\providecommand \EOS [0]{\spacefactor3000\relax}%
\providecommand \BibitemShut  [1]{\csname bibitem#1\endcsname}%
\let\auto@bib@innerbib\@empty
\bibitem [{\citenamefont {Halperin}\ \emph {et~al.}(1993)\citenamefont
  {Halperin}, \citenamefont {Lee},\ and\ \citenamefont {Read}}]{HLR}%
  \BibitemOpen
  \bibfield  {author} {\bibinfo {author} {\bibfnamefont {B.~I.}\ \bibnamefont
  {Halperin}}, \bibinfo {author} {\bibfnamefont {P.~A.}\ \bibnamefont {Lee}}, \
  and\ \bibinfo {author} {\bibfnamefont {N.}~\bibnamefont {Read}},\ }\bibfield
  {title} {\emph {\bibinfo {title} {Theory of the half-filled landau level},\
  }}\href {\doibase 10.1103/PhysRevB.47.7312} {\bibfield  {journal} {\bibinfo
  {journal} {Phys. Rev. B}\ }\textbf {\bibinfo {volume} {47}},\ \bibinfo
  {pages} {7312--7343} (\bibinfo {year} {1993})}\BibitemShut {NoStop}%
\bibitem [{\citenamefont {{Jain}}(2007)}]{JainBook}%
  \BibitemOpen
  \bibfield  {author} {\bibinfo {author} {\bibfnamefont {J.~K.}\ \bibnamefont
  {{Jain}}},\ }\href@noop {} {\emph {\bibinfo {title} {Composite Fermions}}}\
  (\bibinfo  {publisher} {Cambridge University Press},\ \bibinfo {year}
  {2007})\BibitemShut {NoStop}%
\bibitem [{\citenamefont {Kivelson}\ \emph {et~al.}(1997)\citenamefont
  {Kivelson}, \citenamefont {Lee}, \citenamefont {Krotov},\ and\ \citenamefont
  {Gan}}]{Kivelson97}%
  \BibitemOpen
  \bibfield  {author} {\bibinfo {author} {\bibfnamefont {S.~A.}\ \bibnamefont
  {Kivelson}}, \bibinfo {author} {\bibfnamefont {D.-H.}\ \bibnamefont {Lee}},
  \bibinfo {author} {\bibfnamefont {Y.}~\bibnamefont {Krotov}}, \ and\ \bibinfo
  {author} {\bibfnamefont {J.}~\bibnamefont {Gan}},\ }\bibfield  {title} {\emph
  {\bibinfo {title} {Composite-fermion hall conductance at
  \ensuremath{\nu}=1/2},\ }}\href {\doibase 10.1103/PhysRevB.55.15552}
  {\bibfield  {journal} {\bibinfo  {journal} {Phys. Rev. B}\ }\textbf {\bibinfo
  {volume} {55}},\ \bibinfo {pages} {15552--15561} (\bibinfo {year}
  {1997})}\BibitemShut {NoStop}%
\bibitem [{\citenamefont {Barkeshli}\ \emph {et~al.}(2015)\citenamefont
  {Barkeshli}, \citenamefont {Mulligan},\ and\ \citenamefont
  {Fisher}}]{Barkeshli15}%
  \BibitemOpen
  \bibfield  {author} {\bibinfo {author} {\bibfnamefont {M.}~\bibnamefont
  {Barkeshli}}, \bibinfo {author} {\bibfnamefont {M.}~\bibnamefont {Mulligan}},
  \ and\ \bibinfo {author} {\bibfnamefont {M.~P.~A.}\ \bibnamefont {Fisher}},\
  }\bibfield  {title} {\emph {\bibinfo {title} {Particle-hole symmetry and the
  composite fermi liquid},\ }}\href {\doibase 10.1103/PhysRevB.92.165125}
  {\bibfield  {journal} {\bibinfo  {journal} {Phys. Rev. B}\ }\textbf {\bibinfo
  {volume} {92}},\ \bibinfo {pages} {165125} (\bibinfo {year}
  {2015})}\BibitemShut {NoStop}%
\bibitem [{\citenamefont {Geraedts}\ \emph {et~al.}(2016)\citenamefont
  {Geraedts}, \citenamefont {Zaletel}, \citenamefont {Mong}, \citenamefont
  {Metlitski}, \citenamefont {Vishwanath},\ and\ \citenamefont
  {Motrunich}}]{Geraedts15}%
  \BibitemOpen
  \bibfield  {author} {\bibinfo {author} {\bibfnamefont {S.~D.}\ \bibnamefont
  {Geraedts}}, \bibinfo {author} {\bibfnamefont {M.~P.}\ \bibnamefont
  {Zaletel}}, \bibinfo {author} {\bibfnamefont {R.~S.~K.}\ \bibnamefont
  {Mong}}, \bibinfo {author} {\bibfnamefont {M.~A.}\ \bibnamefont {Metlitski}},
  \bibinfo {author} {\bibfnamefont {A.}~\bibnamefont {Vishwanath}}, \ and\
  \bibinfo {author} {\bibfnamefont {O.~I.}\ \bibnamefont {Motrunich}},\
  }\bibfield  {title} {\emph {\bibinfo {title} {The half-filled landau level:
  The case for dirac composite fermions},\ }}\href {\doibase
  10.1126/science.aad4302} {\bibfield  {journal} {\bibinfo  {journal}
  {Science}\ }\textbf {\bibinfo {volume} {352}},\ \bibinfo {pages} {197--201}
  (\bibinfo {year} {2016})}\BibitemShut {NoStop}%
\bibitem [{\citenamefont {Son}(2015)}]{Son15}%
  \BibitemOpen
  \bibfield  {author} {\bibinfo {author} {\bibfnamefont {D.~T.}\ \bibnamefont
  {Son}},\ }\bibfield  {title} {\emph {\bibinfo {title} {Is the composite
  fermion a dirac particle?}\ }}\href {\doibase 10.1103/PhysRevX.5.031027}
  {\bibfield  {journal} {\bibinfo  {journal} {Phys. Rev. X}\ }\textbf {\bibinfo
  {volume} {5}},\ \bibinfo {pages} {031027} (\bibinfo {year}
  {2015})}\BibitemShut {NoStop}%
\bibitem [{\citenamefont {Wang}\ and\ \citenamefont
  {Senthil}(2015)}]{Wang15short}%
  \BibitemOpen
  \bibfield  {author} {\bibinfo {author} {\bibfnamefont {C.}~\bibnamefont
  {Wang}}\ and\ \bibinfo {author} {\bibfnamefont {T.}~\bibnamefont {Senthil}},\
  }\bibfield  {title} {\emph {\bibinfo {title} {Dual dirac liquid on the
  surface of the electron topological insulator},\ }}\href {\doibase
  10.1103/PhysRevX.5.041031} {\bibfield  {journal} {\bibinfo  {journal} {Phys.
  Rev. X}\ }\textbf {\bibinfo {volume} {5}},\ \bibinfo {pages} {041031}
  (\bibinfo {year} {2015})}\BibitemShut {NoStop}%
\bibitem [{\citenamefont {Metlitski}\ and\ \citenamefont
  {Vishwanath}(2015)}]{Metlitski15}%
  \BibitemOpen
  \bibfield  {author} {\bibinfo {author} {\bibfnamefont {M.~A.}\ \bibnamefont
  {Metlitski}}\ and\ \bibinfo {author} {\bibfnamefont {A.}~\bibnamefont
  {Vishwanath}},\ }\bibfield  {title} {\emph {\bibinfo {title} {Particle-vortex
  duality of 2d dirac fermion from electric-magnetic duality of 3d topological
  insulators},\ }}\href@noop {} {\bibfield  {journal} {\bibinfo  {journal}
  {arXiv:1505.05142}\ } (\bibinfo {year} {2015})}\BibitemShut {NoStop}%
\bibitem [{\citenamefont {Wang}\ and\ \citenamefont
  {Senthil}(2016)}]{Wang15rev}%
  \BibitemOpen
  \bibfield  {author} {\bibinfo {author} {\bibfnamefont {C.}~\bibnamefont
  {Wang}}\ and\ \bibinfo {author} {\bibfnamefont {T.}~\bibnamefont {Senthil}},\
  }\bibfield  {title} {\emph {\bibinfo {title} {Half-filled landau level,
  topological insulator surfaces, and three-dimensional quantum spin liquids},\
  }}\href {\doibase 10.1103/PhysRevB.93.085110} {\bibfield  {journal} {\bibinfo
   {journal} {Phys. Rev. B}\ }\textbf {\bibinfo {volume} {93}},\ \bibinfo
  {pages} {085110} (\bibinfo {year} {2016})}\BibitemShut {NoStop}%
\bibitem [{\citenamefont {Metlitski}(2015)}]{MetlitskiSDuality}%
  \BibitemOpen
  \bibfield  {author} {\bibinfo {author} {\bibfnamefont {M.~A.}\ \bibnamefont
  {Metlitski}},\ }\bibfield  {title} {\emph {\bibinfo {title} {{$S$-duality of
  $u(1)$ gauge theory with $\theta =\pi$ on non-orientable manifolds:
  Applications to topological insulators and superconductors}},\ }}\href@noop
  {} {\  (\bibinfo {year} {2015})},\ \Eprint {http://arxiv.org/abs/1510.05663}
  {arXiv:1510.05663 [hep-th]} \BibitemShut {NoStop}%
\bibitem [{\citenamefont {Mross}\ \emph {et~al.}(2015)\citenamefont {Mross},
  \citenamefont {Alicea},\ and\ \citenamefont {Motrunich}}]{Mross15}%
  \BibitemOpen
  \bibfield  {author} {\bibinfo {author} {\bibfnamefont {D.~F.}\ \bibnamefont
  {Mross}}, \bibinfo {author} {\bibfnamefont {J.}~\bibnamefont {Alicea}}, \
  and\ \bibinfo {author} {\bibfnamefont {O.~I.}\ \bibnamefont {Motrunich}},\
  }\bibfield  {title} {\emph {\bibinfo {title} {Explicit derivation of duality
  between a free dirac cone and quantum electrodynamics in (2+1) dimensions},\
  }}\href@noop {} {\bibfield  {journal} {\bibinfo  {journal}
  {arXiv:1510.08455}\ } (\bibinfo {year} {2015})}\BibitemShut {NoStop}%
\bibitem [{\citenamefont {Wang}\ \emph {et~al.}(2014)\citenamefont {Wang},
  \citenamefont {Potter},\ and\ \citenamefont {Senthil}}]{Wang14}%
  \BibitemOpen
  \bibfield  {author} {\bibinfo {author} {\bibfnamefont {C.}~\bibnamefont
  {Wang}}, \bibinfo {author} {\bibfnamefont {A.~C.}\ \bibnamefont {Potter}}, \
  and\ \bibinfo {author} {\bibfnamefont {T.}~\bibnamefont {Senthil}},\
  }\bibfield  {title} {\emph {\bibinfo {title} {Classification of interacting
  electronic topological insulators in three dimensions},\ }}\href {\doibase
  10.1126/science.1243326} {\bibfield  {journal} {\bibinfo  {journal}
  {Science}\ }\textbf {\bibinfo {volume} {343}},\ \bibinfo {pages} {629--631}
  (\bibinfo {year} {2014})}
 \BibitemShut
  {NoStop}%
\bibitem [{\citenamefont {Wang}\ and\ \citenamefont
  {Senthil}(2014)}]{Wang14PRB}%
  \BibitemOpen
  \bibfield  {author} {\bibinfo {author} {\bibfnamefont {C.}~\bibnamefont
  {Wang}}\ and\ \bibinfo {author} {\bibfnamefont {T.}~\bibnamefont {Senthil}},\
  }\bibfield  {title} {\emph {\bibinfo {title} {Interacting fermionic
  topological insulators/superconductors in three dimensions},\ }}\href
  {\doibase 10.1103/PhysRevB.89.195124} {\bibfield  {journal} {\bibinfo
  {journal} {Phys. Rev. B}\ }\textbf {\bibinfo {volume} {89}},\ \bibinfo
  {pages} {195124} (\bibinfo {year} {2014})}\BibitemShut {NoStop}%
\bibitem [{\citenamefont {Halperin}(1982)}]{Halperin82}%
  \BibitemOpen
  \bibfield  {author} {\bibinfo {author} {\bibfnamefont {B.~I.}\ \bibnamefont
  {Halperin}},\ }\bibfield  {title} {\emph {\bibinfo {title} {Quantized hall
  conductance, current-carrying edge states, and the existence of extended
  states in a two-dimensional disordered potential},\ }}\href {\doibase
  10.1103/PhysRevB.25.2185} {\bibfield  {journal} {\bibinfo  {journal} {Phys.
  Rev. B}\ }\textbf {\bibinfo {volume} {25}},\ \bibinfo {pages} {2185--2190}
  (\bibinfo {year} {1982})}\BibitemShut {NoStop}%
\bibitem [{\citenamefont {Ying}\ \emph {et~al.}(1994)\citenamefont {Ying},
  \citenamefont {Bayot}, \citenamefont {Santos},\ and\ \citenamefont
  {Shayegan}}]{Ying94}%
  \BibitemOpen
  \bibfield  {author} {\bibinfo {author} {\bibfnamefont {X.}~\bibnamefont
  {Ying}}, \bibinfo {author} {\bibfnamefont {V.}~\bibnamefont {Bayot}},
  \bibinfo {author} {\bibfnamefont {M.~B.}\ \bibnamefont {Santos}}, \ and\
  \bibinfo {author} {\bibfnamefont {M.}~\bibnamefont {Shayegan}},\ }\bibfield
  {title} {\emph {\bibinfo {title} {Observation of composite-fermion
  thermopower at half-filled landau levels},\ }}\href {\doibase
  10.1103/PhysRevB.50.4969} {\bibfield  {journal} {\bibinfo  {journal} {Phys.
  Rev. B}\ }\textbf {\bibinfo {volume} {50}},\ \bibinfo {pages} {4969(R)}
  (\bibinfo {year} {1994})}\BibitemShut {NoStop}%
\bibitem [{\citenamefont {Bayot}\ \emph {et~al.}(1995)\citenamefont {Bayot},
  \citenamefont {Grivei}, \citenamefont {Manoharan}, \citenamefont {Ying},\
  and\ \citenamefont {Shayegan}}]{Bayot95}%
  \BibitemOpen
  \bibfield  {author} {\bibinfo {author} {\bibfnamefont {V.}~\bibnamefont
  {Bayot}}, \bibinfo {author} {\bibfnamefont {E.}~\bibnamefont {Grivei}},
  \bibinfo {author} {\bibfnamefont {H.~C.}\ \bibnamefont {Manoharan}}, \bibinfo
  {author} {\bibfnamefont {X.}~\bibnamefont {Ying}}, \ and\ \bibinfo {author}
  {\bibfnamefont {M.}~\bibnamefont {Shayegan}},\ }\bibfield  {title} {\emph
  {\bibinfo {title} {Thermopower of composite fermions},\ }}\href {\doibase
  10.1103/PhysRevB.52.R8621} {\bibfield  {journal} {\bibinfo  {journal} {Phys.
  Rev. B}\ }\textbf {\bibinfo {volume} {52}},\ \bibinfo {pages} {R8621--R8624}
  (\bibinfo {year} {1995})}\BibitemShut {NoStop}%
\bibitem [{\citenamefont {Cooper}\ \emph {et~al.}(1997)\citenamefont {Cooper},
  \citenamefont {Halperin},\ and\ \citenamefont {Ruzin}}]{Cooper97}%
  \BibitemOpen
  \bibfield  {author} {\bibinfo {author} {\bibfnamefont {N.~R.}\ \bibnamefont
  {Cooper}}, \bibinfo {author} {\bibfnamefont {B.~I.}\ \bibnamefont
  {Halperin}}, \ and\ \bibinfo {author} {\bibfnamefont {I.~M.}\ \bibnamefont
  {Ruzin}},\ }\bibfield  {title} {\emph {\bibinfo {title} {Thermoelectric
  response of an interacting two-dimensional electron gas in a quantizing
  magnetic field},\ }}\href {\doibase 10.1103/PhysRevB.55.2344} {\bibfield
  {journal} {\bibinfo  {journal} {Phys. Rev. B}\ }\textbf {\bibinfo {volume}
  {55}},\ \bibinfo {pages} {2344--2359} (\bibinfo {year} {1997})}\BibitemShut
  {NoStop}%
\bibitem [{\citenamefont {Ioffe}\ and\ \citenamefont
  {Larkin}(1989)}]{IoffeLarkin}%
  \BibitemOpen
  \bibfield  {author} {\bibinfo {author} {\bibfnamefont {L.~B.}\ \bibnamefont
  {Ioffe}}\ and\ \bibinfo {author} {\bibfnamefont {A.~I.}\ \bibnamefont
  {Larkin}},\ }\bibfield  {title} {\emph {\bibinfo {title} {Gapless fermions
  and gauge fields in dielectrics},\ }}\href {\doibase
  10.1103/PhysRevB.39.8988} {\bibfield  {journal} {\bibinfo  {journal} {Phys.
  Rev. B}\ }\textbf {\bibinfo {volume} {39}},\ \bibinfo {pages} {8988--8999}
  (\bibinfo {year} {1989})}\BibitemShut {NoStop}%
\bibitem [{\citenamefont {Lee}\ and\ \citenamefont {Nagaosa}(1992)}]{Lee92}%
  \BibitemOpen
  \bibfield  {author} {\bibinfo {author} {\bibfnamefont {P.~A.}\ \bibnamefont
  {Lee}}\ and\ \bibinfo {author} {\bibfnamefont {N.}~\bibnamefont {Nagaosa}},\
  }\bibfield  {title} {\emph {\bibinfo {title} {Gauge theory of the normal
  state of high-${\mathit{t}}_{\mathit{c}}$ superconductors},\ }}\href
  {\doibase 10.1103/PhysRevB.46.5621} {\bibfield  {journal} {\bibinfo
  {journal} {Phys. Rev. B}\ }\textbf {\bibinfo {volume} {46}},\ \bibinfo
  {pages} {5621--5639} (\bibinfo {year} {1992})}\BibitemShut {NoStop}%
\bibitem [{\citenamefont {Yokoyama}\ and\ \citenamefont
  {Murakami}(2011)}]{Takehito11}%
  \BibitemOpen
  \bibfield  {author} {\bibinfo {author} {\bibfnamefont {T.}~\bibnamefont
  {Yokoyama}}\ and\ \bibinfo {author} {\bibfnamefont {S.}~\bibnamefont
  {Murakami}},\ }\bibfield  {title} {\emph {\bibinfo {title} {Transverse
  magnetic heat transport on the surface of a topological insulator},\ }}\href
  {\doibase 10.1103/PhysRevB.83.161407} {\bibfield  {journal} {\bibinfo
  {journal} {Phys. Rev. B}\ }\textbf {\bibinfo {volume} {83}},\ \bibinfo
  {pages} {161407} (\bibinfo {year} {2011})}\BibitemShut {NoStop}%
\bibitem [{\citenamefont {Nagaosa}\ \emph {et~al.}(2010)\citenamefont
  {Nagaosa}, \citenamefont {Sinova}, \citenamefont {Onoda}, \citenamefont
  {MacDonald},\ and\ \citenamefont {Ong}}]{AHERMP}%
  \BibitemOpen
  \bibfield  {author} {\bibinfo {author} {\bibfnamefont {N.}~\bibnamefont
  {Nagaosa}}, \bibinfo {author} {\bibfnamefont {J.}~\bibnamefont {Sinova}},
  \bibinfo {author} {\bibfnamefont {S.}~\bibnamefont {Onoda}}, \bibinfo
  {author} {\bibfnamefont {A.~H.}\ \bibnamefont {MacDonald}}, \ and\ \bibinfo
  {author} {\bibfnamefont {N.~P.}\ \bibnamefont {Ong}},\ }\bibfield  {title}
  {\emph {\bibinfo {title} {Anomalous hall effect},\ }}\href {\doibase
  10.1103/RevModPhys.82.1539} {\bibfield  {journal} {\bibinfo  {journal} {Rev.
  Mod. Phys.}\ }\textbf {\bibinfo {volume} {82}},\ \bibinfo {pages}
  {1539--1592} (\bibinfo {year} {2010})}\BibitemShut {NoStop}%
\bibitem [{\citenamefont {Xiao}\ \emph {et~al.}(2006)\citenamefont {Xiao},
  \citenamefont {Yao}, \citenamefont {Fang},\ and\ \citenamefont
  {Niu}}]{Xiao06}%
  \BibitemOpen
  \bibfield  {author} {\bibinfo {author} {\bibfnamefont {D.}~\bibnamefont
  {Xiao}}, \bibinfo {author} {\bibfnamefont {Y.}~\bibnamefont {Yao}}, \bibinfo
  {author} {\bibfnamefont {Z.}~\bibnamefont {Fang}}, \ and\ \bibinfo {author}
  {\bibfnamefont {Q.}~\bibnamefont {Niu}},\ }\bibfield  {title} {\emph
  {\bibinfo {title} {Berry-phase effect in anomalous thermoelectric
  transport},\ }}\href {\doibase 10.1103/PhysRevLett.97.026603} {\bibfield
  {journal} {\bibinfo  {journal} {Phys. Rev. Lett.}\ }\textbf {\bibinfo
  {volume} {97}},\ \bibinfo {pages} {026603} (\bibinfo {year}
  {2006})}\BibitemShut {NoStop}%
\bibitem [{\citenamefont {Qin}\ \emph {et~al.}(2011)\citenamefont {Qin},
  \citenamefont {Niu},\ and\ \citenamefont {Shi}}]{Qin11}%
  \BibitemOpen
  \bibfield  {author} {\bibinfo {author} {\bibfnamefont {T.}~\bibnamefont
  {Qin}}, \bibinfo {author} {\bibfnamefont {Q.}~\bibnamefont {Niu}}, \ and\
  \bibinfo {author} {\bibfnamefont {J.}~\bibnamefont {Shi}},\ }\bibfield
  {title} {\emph {\bibinfo {title} {Energy magnetization and the thermal hall
  effect},\ }}\href {\doibase 10.1103/PhysRevLett.107.236601} {\bibfield
  {journal} {\bibinfo  {journal} {Phys. Rev. Lett.}\ }\textbf {\bibinfo
  {volume} {107}},\ \bibinfo {pages} {236601} (\bibinfo {year}
  {2011})}\BibitemShut {NoStop}%
\bibitem [{\citenamefont {Luttinger}(1964)}]{Luttinger64}%
  \BibitemOpen
  \bibfield  {author} {\bibinfo {author} {\bibfnamefont {J.~M.}\ \bibnamefont
  {Luttinger}},\ }\bibfield  {title} {\emph {\bibinfo {title} {Theory of
  thermal transport coefficients},\ }}\href {\doibase
  10.1103/PhysRev.135.A1505} {\bibfield  {journal} {\bibinfo  {journal} {Phys.
  Rev.}\ }\textbf {\bibinfo {volume} {135}},\ \bibinfo {pages} {A1505--A1514}
  (\bibinfo {year} {1964})}\BibitemShut {NoStop}%
\bibitem [{\citenamefont {Altshuler}\ and\ \citenamefont
  {Ioffe}(1992)}]{Altshuler92}%
  \BibitemOpen
  \bibfield  {author} {\bibinfo {author} {\bibfnamefont {B.~L.}\ \bibnamefont
  {Altshuler}}\ and\ \bibinfo {author} {\bibfnamefont {L.~B.}\ \bibnamefont
  {Ioffe}},\ }\bibfield  {title} {\emph {\bibinfo {title} {Motion of fast
  particles in strongly fluctuating magnetic fields},\ }}\href {\doibase
  10.1103/PhysRevLett.69.2979} {\bibfield  {journal} {\bibinfo  {journal}
  {Phys. Rev. Lett.}\ }\textbf {\bibinfo {volume} {69}},\ \bibinfo {pages}
  {2979--2982} (\bibinfo {year} {1992})}\BibitemShut {NoStop}%
\bibitem [{\citenamefont {Mirlin}\ \emph {et~al.}(1998)\citenamefont {Mirlin},
  \citenamefont {Polyakov},\ and\ \citenamefont {W\"olfle}}]{Mirlin98}%
  \BibitemOpen
  \bibfield  {author} {\bibinfo {author} {\bibfnamefont {A.~D.}\ \bibnamefont
  {Mirlin}}, \bibinfo {author} {\bibfnamefont {D.~G.}\ \bibnamefont
  {Polyakov}}, \ and\ \bibinfo {author} {\bibfnamefont {P.}~\bibnamefont
  {W\"olfle}},\ }\bibfield  {title} {\emph {\bibinfo {title} {Composite
  fermions in a long-range random magnetic field: Quantum hall effect versus
  {S}hubnikov de {H}aas oscillations},\ }}\href {\doibase
  10.1103/PhysRevLett.80.2429} {\bibfield  {journal} {\bibinfo  {journal}
  {Phys. Rev. Lett.}\ }\textbf {\bibinfo {volume} {80}},\ \bibinfo {pages}
  {2429--2432} (\bibinfo {year} {1998})}\BibitemShut {NoStop}%
\end{thebibliography}
%

\end{document}